\documentclass[11pt,a4paper]{article}

\usepackage{amssymb}
\usepackage{amsmath}
\usepackage{amsfonts}
\usepackage{graphicx, float}

\usepackage{color}
\usepackage{enumerate}
\usepackage[american]{babel}

\usepackage{hyperref}
\hypersetup{
	pdftitle={Unparticle contribution to the hydrogen atom ground state energy},
	colorlinks,
    linkcolor={red!50!black},
    citecolor={blue!50!black},
    urlcolor={blue!80!black}
}

\usepackage[draft,colorinlistoftodos,shadow,textsize=small]{todonotes}   
\presetkeys{todonotes}{noline}{}

\usepackage{longtable}
\usepackage{booktabs}
\usepackage{epstopdf}
\usepackage{textcomp,gensymb}
\newcommand{\Mpl}{M_{\rm Pl}}
\newcommand{\tabheadfont}[1]{\textbf{#1}}
\newcommand{\un}{\ensuremath{\mathcal{U}}}
\newcommand{\bz}{\ensuremath{\mathcal{BZ}}}
\newcommand{\du}{\ensuremath{d_\mathcal{U}}}
\newcommand{\lu}{\ensuremath{\Lambda_\mathcal{U}}}
\newcommand{\mun}{\ensuremath{M_\mathcal{U}}}
\newcommand{\gam}[1]{\ensuremath{\Gamma \! \left(#1\right)}}
\newcommand{\lae}{\ensuremath{\lambda_\mathrm{e}}}
\newcommand{\lap}{\ensuremath{\lambda_\mathrm{p}}}
\newcommand{\indicesE}[2]{\ensuremath{E_\mathrm{#1}^\mathrm{#2}}}
\newcommand{\indexrm}[1]{\ensuremath{\mathrm{#1}}}

\usepackage[nosort]{cite}
\usepackage{color,colordvi}

\begin{document}

\begin{center}
{\Large\bf 
Unparticle contribution to the\\hydrogen atom ground state energy}  
\end{center}

\vspace{-0.1cm}

\begin{center}
{\bf 
Michael F. Wondrak}$^{a,b}$\footnote{wondrak@fias.uni-frankfurt.de},
{\bf Piero Nicolini}$^{a,b}$\footnote{nicolini@fias.uni-frankfurt.de}
{\bf and Marcus Bleicher}$^{a,b}$\footnote{bleicher@fias.uni-frankfurt.de}

\vspace{.6truecm}


{\em $^a$Frankfurt Institute for Advandced Studies (FIAS),\\
Ruth-Moufang-Str.~1, 60438 Frankfurt am Main, Germany}\\

\vspace{.2truecm}
{\em $^b$Institut f\"ur Theoretische Physik,\\
Johann Wolfgang Goethe-Universit\"at Frankfurt am Main,\\Max-von-Laue-Str.~1,
60438 Frankfurt am Main, Germany}\\
\vspace{.6truecm}
\today 
\end{center}

\vspace{0.1cm}

\begin{abstract}
\noindent  
 {\small In the present work we  study the effect of unparticle modified static potentials on the energy levels of the hydrogen atom.  By using Rayleigh-Schr\"odinger perturbation theory, we obtain the energy shift of the ground state  and we compare it with experimental data. Bounds on the unparticle energy scale \lu{} as a function of the scaling dimension \du{} and the coupling constant $\lambda$ are derived. We show that there exists a parameter region where bounds on \lu{} are stringent, signalling that unparticles could be tested in atomic physics experiments.
 

 } 
\end{abstract}

\thispagestyle{empty}
\clearpage
\section{Introduction}

Unparticle physics is an extension of the Standard Model, consisting in the possibility of having a non-trivial scale invariant, yet undiscovered sector of particle physics. 

At first sight, unparticles appear as a generalization of neutrinos because they share the following properties: scale invariance and only very weak interaction with other fields. Neutrinos enjoy the first property to a good approximation, although their oscillations disclose a small non-zero mass. The second property is a general requirement of any hypothetical particle sector because we want it to be, to a certain extent, hidden from current observations. On a closer inspection, however, we find that unparticles differ drastically from neutrinos. Since we do not restrict the unparticle fields to be massless, we can no longer speak in terms of particle number as in the conventional manner. The unparticle field is controlled by a canonical scaling dimension \du{} which is in general a non-integer number. Due to the unusual character, one refers to the matter described by such a theory as \textit{unlike particles}, or unparticle stuff.

After  Georgi's seminal paper  \cite{Georgi2007a} unparticle effects have been explored in many areas spanning collider physics \cite{Georgi2007b,CheungKY2007a,CheungKY2007b,CMS2015a,
CMS2015b,KathreinKS2011,Liao2007},  gauge and Higgs interactions \cite{FoxRS2007,GaS14}, cosmology and astrophysics \cite{Davoudiasl2007,DasMR2008}, AdS/CFT correspondence \cite{CacciapagliaMT2008}  as well as gravity short scale deviations \cite{GoldbergN2008,GaeteS2008} and black holes \cite{Mureika2008,Mureika2009,GHS10,MureikaS2010,Mur12a}.
Unparticles also play a crucial role in the fractal properties of a quantum spacetime. A new fractality indicator, called un-spectral dimension, has recently been  proposed to address the case of a random-walker problem in terms of an unparticle probe \cite{NiS11}. When the manifold topological dimension is $2$, the un-spectral dimension turns to be $2\du$, \textit{i.e.}, it depends only on the scaling dimension \du{}. This fact explains the complete ``fractalizazion´´ of the event horizon of un-gravity  black holes \cite{GHS10,MureikaS2010}, as well as of metallic plates for the Casimir effect   in the presence of an un-photon field \cite{FrassinoNP2014}. Finally, unparticles have been proposed to explain some anomalies in currents flowing in super-conductors \cite{LeBlancG2015} and transport phenomena in cuprates \cite{KLP15}.

In this paper we want to address one of the basic questions of unparticle physics, \textit{i.e.}, the value of $\Lambda_{\cal U}$, the typical energy scale of the theory. To achieve this goal, we consider the modifications of static potentials that emerge from virtual unparticle exchange. Specifically by considering the corrections to the Coulomb potential we calculate the deviations of the ground state energy of the hydrogen atom. We show that competitive bounds on  $\Lambda_{\cal U}$ can be derived by a comparison with experimental data. 

The paper is organized as follows. After a short review of the basic formalism of unparticle physics (Section \ref{sec:unparticle}), we present the calculation of the energy shift  by a perturbative solution of the Schr\"{o}dinger equation in the presence of an unparticle modified electrostatic potential (Section \ref{sec:hatom}).  Finally, in Section \ref{sec:concl} we draw our conclusions.

\section{Unparticle physics and static potentials}
\label{sec:unparticle}

We recap the basic motivations of  unparticle physics along the lines of Georgi \cite{Georgi2007a}. We start by considering that at some very high energy scale, the Standard Model is accompanied by an additional sector of  Banks-Zaks fields $\left(\bz\right)$. The interaction between the two sectors takes place by exchange of mediating particles having  a large mass scale \mun{}. If our energy scale of interest falls below \mun{}, we can apply effective field theory to integrate out the mediating field \cite{GrinsteinIR2008} and get the final interaction Lagrangian
\begin{equation}
\frac{1}{\mun^k} \, \mathcal{O}_\indexrm{SM} \, \mathcal{O}_\bz
\label{eq:ThBackground_CouplingBZSM_Def}
\end{equation}
where $\mathcal{O}_\indexrm{SM}$ denotes a Standard Model field operator of scaling dimension $d_\indexrm{SM}$ and $\mathcal{O}_\bz$ is a Banks-Zaks field operator of scaling dimension $d_\bz$. The factor  $\mun^{-k}$ guarantees the dimensional consistency of \eqref{eq:ThBackground_CouplingBZSM_Def}, being $k=d_\indexrm{SM}+d_\bz-D$ and $D$ is the spacetime dimension.

If the energy is further decreased to a scale $\lu{}<\mun$, the Banks-Zaks fields undergo a dimensional transmutation and exhibit a scale invariant behavior with a continuous mass distribution. For energies lower than \lu{}, the \bz{} sector becomes unparticle operators $\mathcal{O}_\un$. The matching conditions onto the Banks-Zaks operators are imposed at the energy scale \lu{} and determine the structure of the coupling between the Standard Model and the unparticle fields based on (\ref{eq:ThBackground_CouplingBZSM_Def}):
\begin{equation}
\frac{C_{\un}\,\lu^{d_{\bz}-d_\un}}{\mun^k} \, \mathcal{O}_\indexrm{SM} \, \mathcal{O}_\un
= \frac{\lambda}{\lu^{d_\indexrm{SM}+d_\un-D}} \, \mathcal{O}_\indexrm{SM} \, \mathcal{O}_\un
\label{eq:ThBackground_CouplingUSM_Def}
\end{equation}
where $d_\un$ is the scaling dimension of the unparticle operator $\mathcal{O}_\un$, $C_\un$ denotes a dimensionless constant  and $\lambda$ is a dimensionless coupling parameter defined by
\begin{equation}
\lambda = C_\un\,{\left(\frac{\lu}{\mun}\right)}^k < 1 \, .
\label{eq:ThBackground_lambda_Def}
\end{equation}
The inequality holds if $C_\un<1$ and $d_\bz>D-d_\indexrm{SM}$. 
Any experimental bound on the interaction allows for constraints on the unparticle parameter space, \textit{i.e.}, \lu{}, $\lambda$ and \du. The scale hierarchy is $1\ \mathrm{TeV}\leq\lu<\mun\leq\Mpl$, where $\Mpl$ is the Planck mass \cite{Mur12a}. In the rest of the present paper we assume $D=4$. We also assume the customary interval $1<\du<2$ \cite{Georgi2007b}. The case $\du=1$ corresponds to no fractalizazion or other continuous dimension effects of unparticle physics. 

Unparticles have been largely employed in context of static potential emerging from virtual particle exchange \cite{GaS14,GoldbergN2008,GaeteS2008,DeshpandeHJ2008,MuS10}. Such results are instrumental to the working hypothesis of the current investigation. In view of the analysis of the hydrogen atom, we consider an additional contribution to the Coulomb potential for the presence of unparticle exchange in the interaction between electron and proton. 
\begin{figure}[ht]
  \centering
  \includegraphics[width=0.4\linewidth]{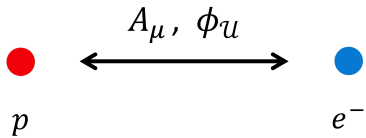}
  \caption{Electron-proton interaction due to a vector field $A_\mu$ and an unparticle scalar field $\phi_{\cal U}$.}
  \label{fig:interaction}
\end{figure}
In general, we assume electrons and protons to carry unparticle charges $\lae $ and $\lap$, respectively. At the same time, they also possess electric charges $\pm e$  allowing them to couple to the photon field $A_\mu$. Both interactions are independent of each other since we assume the effective couplings between unparticle stuff and photons to be negligible. Therefore the interaction Lagrangian can be written as
\begin{equation}
{\cal L}_{\mathrm{int}}=J^{\mu} A_\mu + \frac{1}{(\lu)^{\du-1}}\ J_{\cal U} \phi_{\cal U}
\label{eq:ThBackground_L_int_scalar}
\end{equation}
where $J^{\mu}=j(\vec{x})\ \delta^\mu_0$ and $J_{\cal U}=j_{\cal U}(\vec{x})$ and
\begin{eqnarray}
&&j(\vec{x})=- e\ \delta(\vec{x}-\vec{x}_{\rm e})+e\ \delta(\vec{x}-\vec{x}_{\rm p})\\
&&j_{\cal U}(\vec{x})= \lae\ \delta(\vec{x}-\vec{x}_{\rm e})+\lap\ \delta(\vec{x}-\vec{x}_{\rm p}).
\end{eqnarray}   
Alternatively one can consider the vector unparticle interaction Lagrangian\footnote[1]{The Lagrangian in such a case reads 
\begin{equation*}
{\cal L}_{\mathrm{int}}=J^{\mu} A_\mu + \frac{1}{(\lu)^{\du-1}}\ J^{\mu}_{\cal U} (A_{\cal U})_{ \mu}
\end{equation*} 
where  $J^{\mu}_{\cal U}=j_{\cal U}(\vec{x})\ \delta^\mu_0$.
} and derive the static potential much in the same way as in the scalar case. The two results do not differ apart from a global sign (see \textit{e.g.}\cite{GaeteS2008}). We recall, however, that conformal invariance can be lost for vector fields if $\du<3$, although pure scale invariance can be preserved. For scalar unparticles the issue does not arise \cite{GrinsteinIR2008}.

The expression for the unparticle interaction energy $V_{\cal U}$ between an electron and a proton in the static case reads \cite{GaS14,GoldbergN2008,GaeteS2008,DeshpandeHJ2008,MuS10}: 
\begin{equation}
V_{\cal U}
= -\xi_{\du} \, \left(\frac{\lae\,\lap}{\lu^{2\du-2} \, r^{2\du-1}}\right)
\label{eq:IntEnergyUn_concreteExpr_scalar_final}
\end{equation}
where the coefficient is
$\xi_{\du}
\equiv \dfrac{\sqrt{\pi}}{{\left(2\pi\right)}^{2\du}} \, 
       \dfrac{\gam{\du-\frac{1}{2}}}{\gam{\du}}
$
and $r\equiv|\vec{x}_{\rm e}-\vec{x}_{\rm p}|$ denotes the distance between the charges.

\begin{figure}[!ht]
  \centering
  \includegraphics[width=\linewidth]{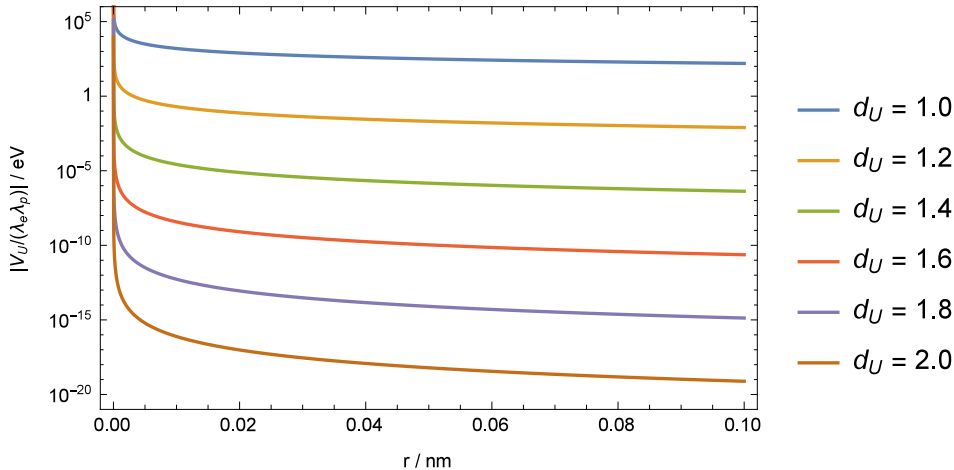}
  \caption{Interaction energy per unparticle charge unit vs.~$r$ for different values of $\du$ in the case $\lu=10\,\mathrm{TeV}$.}
  \label{fig:V_vs_r-2D-dU1until2-legend}
\end{figure}

For the typical size of the hydrogen atom we can estimate the energy correction with the help of Fig.~\ref{fig:V_vs_r-2D-dU1until2-legend}. For $r=0.05\,\mathrm{nm}$ the energy shift per unit unparticle charge varies between $3\times 10^2\,\mathrm{eV}$ and $6\times 10^{-19}\, \mathrm{eV}$ in the range $1<\du<2$ with $\lu=10\,\mathrm{TeV}$.

\section{Unparticle effects  in the hydrogen atom}
\label{sec:hatom}
The hydrogen atom is a well investigated system. 
In the non-relativistic limit we can apply the Schr\"odinger formalism to describe electron dynamics in a static Coulomb potential of the proton, by using  the particle reduced mass $\mu$.
The energy spectrum in Gaussian/natural units ($4\pi\epsilon_0=\hbar=c=1$) reads
\begin{equation}
E_n 
= -\frac{\mu \, \alpha^2}{2 n^2}
\label{eq:ThBackground_Schrod_ESpectrum}
\end{equation}
for all $n \in \mathbb{N}$. Here we used the fine-structure constant $\alpha \equiv  e^2$. 
The biggest corrections to the Schr\"odinger description are of order $\alpha^4$ and arise from relativistic effects  (\textit{i.e.} kinetic, spin-orbit and Darwin terms) that can be included via fine-structure modifications of the Hamiltonian. 

There exist higher order corrections as well. The proton possesses a finite size which affects the form of the Coulomb potential for distances shorter than its radius. Since the proton has a spin which interacts with both the electron angular motion and the electron spin, hyperfine corrections arise.  Finally, one can include QED corrections such as  the Lamb shift.
In Tab.~\ref{tab:ThBackground_Energy_Comparison}, the ground state energy in different descriptions is displayed. Note that the uncertainty of the  two most accurate values arises from the uncertainty in the Planck constant $h=4.135\,667\,662\,\left(25\right) \times {10}^{-15} \, \mathrm{eVs}$ \cite{MohrNT2015}.
\begin{table}[!ht]
\centering
\begin{tabular}[l]{p{0.11\textwidth}p{0.375\textwidth}p{0.515\textwidth-6\tabcolsep}}
  \tabheadfont{Energy}&\tabheadfont{Description}&\tabheadfont{Value}\\
  \midrule
  $\indicesE{th}{S}$
  &Schr\"odinger, non-relativistic& $-13.598\,286\,\mathrm{eV}$\\[3pt]
  $\indicesE{th}{S,\,rel}$
  &Schr\"odinger, & $-13.598\,467\,\mathrm{eV}$\\
  &incl.~fine-structure&\\[3pt]
  $\indicesE{th}{D}$
  &Dirac& $-13.598\,467\,\mathrm{eV}$\\[3pt]
  $\indicesE{th}{QED}$
  &currently best theor. value& $-13.598\,434\,49\,\left(9\right)\, \mathrm{eV}$\\
  &\!\!\cite{JentschuraKLMT2005}& $3\,288\,086\,857.127\,6\,\left(3\,1\right)\, \mathrm{MHz} \cdot h$ \\[3pt]
  $E_\indexrm{exp}$
  &currently best exp.~value&  $-13.598\,434\,48\,\left(9\right)\, \mathrm{eV}$\\
  &\!\!\cite{Kramida2010}&$3\,288\,086\,856.8\,\left(0.7\right)\, \mathrm{MHz} \cdot h$\\[3pt]
\end{tabular}
\caption{Theoretical and experimental values of the hydrogen ground state energy.}
\label{tab:ThBackground_Energy_Comparison}
\end{table}

We start our analysis by considering the energy level shift due to unparticle effects within the non-relativistic description. The radial Schr\"odinger equation of the hydrogen atom in the presence of both electrostatic and unparticle potentials reads
\begin{equation}
\left( 
  -\frac{1}{2\mu} \, \frac{{\rm d}^2}{{\rm d}r^2}
  +\frac{1}{2\mu} \, \frac{l \left(l+1\right)}{r^2}
  -\frac{e^2}{r}
  -\xi_{\du} \, \frac{\lae\,\lap}{\lu^{2\du-2} \, r^{2\du-1}}
\right)
u_{nl}\!\left(r\right) 
= E_{nlm}\,u_{nl}\!\left(r\right) 
\label{eq:SchEq}
\end{equation}
where the complete wavefunction reads
\[\psi_{nlm}\!\left(r,\,\vartheta,\,\varphi\right)
= \frac{u_{nl}\left(r\right)}{r}\,Y_{lm}\!\left(\vartheta,\,\varphi\right).\]
The above equation is a second order, linear, ordinary differential equation with non-polynomial coefficients. To our best knowledge no exact solution is available in the literature but one can still rely on  perturbation theory. This is justified since $\lu>1$ TeV by hypothesis, \textit{i.e.}, unparticle physics can lead only to subleading corrections to the Standard Model. 
As a result the  energy shift of the ground state at the first order in  perturbation theory reads:
\begin{eqnarray}
\Delta_{100}^{\left(1\right)}
&=& \left\langle {100}^{\left(0\right)} \,\right|\, 
    -\xi_{\du} \, \frac{\lae\,\lap}{\lu^{2\du-2} \, r^{2\du-1}} 
    \,\left|\, {100}^{\left(0\right)} \right\rangle \nonumber\\
&=& -4\, \xi_{\du} \, \frac{\lae\,\lap}{\lu^{2\du-2} \, a^3}
    \int_0^\infty\!{\rm d}r \; r^{-2\du+3} \, e^{\frac{-2r}{a}} 
    \nonumber\\
&=& -\frac{1}{{\left(2\pi\right)}^{2\du-2}} \, 
    \frac{\left(\du-1\right) \, \left(\frac{3}{2}-\du\right)}
    {\sin\!\left(2\pi\du\right) \, {\left(\gam{\du}\right)}^2} \,
    \frac{\lae\,\lap}{\lu^{2\du-2} \, a^{2\du-1}}\, ,
\label{eq:PertTh_EnShift_1stOrd}
\end{eqnarray}
where we used $\psi_{100}^{(0)}
= \frac{2\, }{\sqrt{4\pi a^3}} \, e^{-\frac{r}{a}}
$ with $a\equiv 1/(\alpha\mu)\simeq 5.29\times 10^{-11}$~m. 

For $1< \du < 2$, the sign of $\Delta_{100}^{\left(1\right)}$ is determined by the product of \lae{} and \lap{}. This means that like unparticle charges lead to a shift to lower energies while unlike charges cause a shift to higher energies.

The theoretical energy value $E_{\rm th}$ can be divided up into the Schr\"odinger contribution $\indicesE{th}{S}$ and higher order terms $\indicesE{th}{HO}$. 
The same applies to the unparticle-modified quantities. As a result we can approximate the energy shift as
\begin{eqnarray}
\left| E_{\rm th,\,\un} - E_{\rm th} \right| &=& \left| \left(\indicesE{th,\,\un}{S}+\indicesE{th,\,\un}{HO}\right) 
- \left(\indicesE{th}{S}+\indicesE{th}{HO}\right) \right| \nonumber\\
&\simeq & \left| \indicesE{th,\,\un}{S} - \indicesE{th}{S} \right|.
\label{eq:Constr_UnpContr_S}
\end{eqnarray}
Here $\indicesE{th,\,\un}{HO}$ stands for all the higher order contributions in the presence of unparticles, contributions we deliberately neglected in our first order description. This assumption introduces a theoretical error $\delta E_{\rm th} \sim \left|\indicesE{th,\,\un}{HO}-\indicesE{th}{HO}\right|$. Thus the bound on the energy shift can be written as
\begin{equation}
\left| \indicesE{th,\,\un}{S} - \indicesE{th}{S} \right|=\left| \Delta_{100}^{\left(1\right)} \right|<\delta E_{\rm th}+\delta E_{\rm exp}\, ,
\end{equation}
where $\delta E_{\rm exp}$ is the experimental error. 
From Tab. \ref{tab:ThBackground_Energy_Comparison}, one can see that the Schr\"{o}dinger description differs from the QED result by about $1.5\times 10^{-4}$ eV. The theoretical relative error can be estimated to be $\delta E_{\rm th}/\left|\indicesE{th}{S}\right| \simeq 1.1\times 10^{-5}$. It dominates over the experimental relative error that is of the order of $2.2 \times 10^{-10}$  \cite{Kramida2010}.  Accordingly we define where $\delta_{\mathrm{max}}\equiv (\delta E_{\rm th}+\delta E_{\rm exp})/\indicesE{th}{S}\simeq 1.1\times 10^{-5}$ to get 
\begin{equation}
\left|\frac{\Delta_{100}^{\left(1\right)}}{\indicesE{th}{S}}\right|
= \frac{2\alpha^{2\du-3}\,\mu^{2\du-2}}{{\left(2\pi\right)}^{2\du-2}} \, 
    \frac{\left(\du-1\right) \, \left(\frac{3}{2}-\du\right)}
    {\sin\!\left(2\du\pi\right) \, {\left(\gam{\du}\right)}^2} \,
    \frac{\left| \lae\,\lap \right|}{\lu^{2\du-2}}
< \delta_{\mathrm{max}}\,.
\label{eq:Constr_AnalytTerm}
\end{equation}
From \eqref{eq:Constr_AnalytTerm} we obtain
\begin{equation}
\lu 
\geq \frac{\alpha\,\mu}{2\pi} \, {\left( \frac{2}{\alpha} \, \frac{\left(\du-1\right) \left(\frac{3}{2}-\du\right)}{\sin\!\left(2\pi \du\right) {\left(\gam{\du}\right)}^2} \, \frac{\left| \lae\,\lap \right|}{\delta_{\mathrm{max}}} \right)}^{\frac{1}{2\du-2}}\, .
\label{eq:constraintHAtom}
\end{equation}
 \begin{figure}[!ht]
  \centering
  \includegraphics[width=1\linewidth]{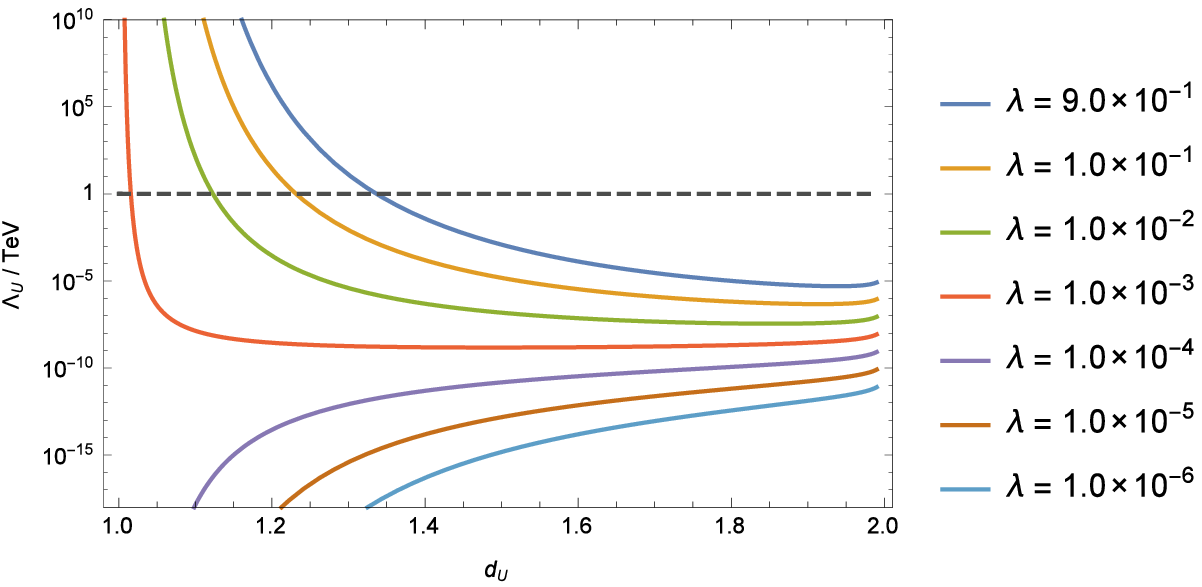}
  \caption{Lower bounds for $\lu$ with respect to $\du$ and $\lambda$.}
  \label{fig:LU_vs_dU-2D-lambdaE-1untilE-6-legend}
\end{figure}
In Fig.~\ref{fig:LU_vs_dU-2D-lambdaE-1untilE-6-legend} the lower bounds for \lu{} are illustrated for different values of the unparticle charge $\lambda$, where we assumed $|\lae|=|\lap|\equiv\lambda$. The area above each curve is the allowed parameter space for the chosen value of $\lambda$. 
If the parameter $\lambda$ is larger than a threshold value, $\lambda > \lambda_{\rm th} \equiv \sqrt{2\pi\alpha\,\delta_{\rm max}} \simeq 7.1\times 10^{-4}$, the lower bound on \lu{} is diverging for $\du\to 1$ and exceeds $1$ TeV in some parameter region. Here $\lambda_{\rm th}$ is obtained by requiring that the base of the exponentiation in \eqref{eq:constraintHAtom} equal $1$.   
The limit $\lu \to \infty$ corresponds to vanishing unparticle effects since the unparticle sector in the Lagrangian \eqref{eq:ThBackground_L_int_scalar} disappears yielding the standard electrodynamics limit ${\cal L}_{\mathrm{int}}\to J^{\mu} A_\mu$.
 This explains the divergent lower bound on \lu{} expected also in other experiments such as the electron $g-2$ anomaly \cite{Liao2007} or the Casimir effect \cite{FrassinoNP2014}.
In contrast, for $\lambda\leq\lambda_{\rm th}$, the scale \lu{} is weakly
constrained in the limit $\du\to 1$. This  means that the unparticle
contribution lies within $\delta_{\rm max}$ irrespective of the value of \lu.

\section{Discussion and conclusions}
\label{sec:concl}

In the current work, we have derived the corrections to the hydrogen atom ground state energy due to the presence of an unparticle modified static potential. We obtained this result within a perturbative analysis of the non-relativistic theory of the hydrogen atom. Our result lets obtain compelling limits on the unparticle scale \lu. For $\du\lesssim 1.3$, bounds on \lu{} exceed $1$ TeV. Contrary to other proposed investigations (\textit{e.g.} Newton's law correction \cite{GoldbergN2008}, proton-proton collisions at the LHC \cite{CMS2015a,CMS2015b}), the presented analysis can capture the key feature of unparticle physics, \textit{i.e.}, the dependence on the continuous scaling dimension $\du$. 

The presented results are filling a gap in the literature and are opening the route for further investigations. For instance one may be interested to find an exact solution of the Schr\"{o}dinger equation \eqref{eq:SchEq} emerging from the inclusion of unparticle static potentials.  Higher order corrections to the non-relativistic description can also be included in the analysis in order to approach the bounds currently offered by $g-2$ analyses \cite{Liao2007}. 

From the current non-relativistic analysis one can draw an important conclusion: Unparticle effects might be tested in atomic physics experiments.

\subsection*{Acknowledgements}
 
This work has been supported by the Studienstiftung des deutschen Volkes and 
by the Helmholtz International Center for FAIR
within the framework of the LOEWE program (Landesoffensive zur Entwicklung Wissenschaftlich- \"{O}konomischer Exzellenz) launched by the State of Hesse.

\providecommand{\href}[2]{#2}\begingroup\raggedright

\endgroup

\end{document}